\documentclass[aps,prapplied,preprint,12pt,groupedaddress,notitlepage]{revtex4-2}  
\usepackage[utf8]{inputenc}
\usepackage[T1]{fontenc}
\usepackage{amsmath,amssymb,graphicx,wrapfig,setspace,placeins,csquotes,booktabs}
\begin{document}
\preprint{APS/123-QED}
\title{Directional Thermal Emission Across Both Polarizations in Planar Photonic Architectures}

\author{David E. Abraham}
\author{Daniel Cui}
\author{Baolai Liang}
\author{Jae S. Hwang}
\author{Parthiban Santhanam}
\author{Linus Kim}
\author{Rayen Lin}
\author{Aaswath P. Raman}
\email{aaswath@ucla.edu}

\affiliation{Department of Materials Science and Engineering, University of California, Los Angeles, Los Angeles, CA 90095, USA}
\affiliation{California NanoSystems Institute, University of California, Los Angeles, Los Angeles, CA 90095, USA}

\begin{abstract}

Directional and spectral control of thermal emission is essential for applications in energy conversion, imaging, and sensing. Existing planar, lithography-free epsilon-near-zero (ENZ) films only support transverse-magnetic (TM) control of thermal emission via the Berreman mode and cannot address transverse-electric (TE) waves due to the absence of natural optical magnetism over optical and infrared wavelengths Here, we introduce a hyperbolic metamaterial comprising alternating layers of degenerately-doped and intrinsic InAs that exhibits an epsilon-and-mu-near-zero (EMNZ) response, enabling dual-polarized, directionally and spectrally selective thermal emission. We first theoretically demonstrate that a mu-near-zero (MNZ) film on a perfect magnetic conductor supports a “magnetic” Berreman mode, absorbing TE-polarized radiation in analogy to the conventional Berreman mode supported in TM polarization. Using genetic and gradient-descent optimization, we design a dual-polarized emitter with independently tunable spectral peaks and emission angles. Parameter retrieval via homogenization confirms simultaneous EMNZ points at the target wavelengths and angles. Finally, experimental measurement of a sample fabricated via molecular beam epitaxy exhibits high absorptivity peaks for both polarizations in close agreement with simulations. This work realizes lithography-free, dual-polarized, spectrally and directionally selective emitters, offering a versatile platform for advanced infrared thermal management and device integration.

\end{abstract}

\maketitle

\section{Introduction}

The ability to modify thermal emission both spectrally and directionally is of great importance to a wide range of technologies in energy conversion, imaging, and sensing applications. To that end, a broad variety of photonic strategies.\cite{hesketh_organ_1986,enoch_metamaterial_2002, wang_ultra-broadband_2024, johns_tailoring_2022}
However, tailoring the directionality of thermal emission has proven to be more challenging, often requiring lithography \cite{fan_directional_2024} to introduce gratings or metallic patterning, or relying on thicker films to produce Fabry-Perot resonances \cite{sarkar_lithography-free_2024}.

A well-studied strategy for achieving directional thermal emission without the use of lithography uses deep-subwavelength films of epsilon near zero (ENZ) media on metallic substrates. These systems are known to support a leaky surface mode, known as the Berreman mode, which leads to a near-grazing angle of incidence absorption peak near the ENZ frequency of the film. \cite{vassant_berreman_2012, chen_trapping_2013, liberal_near-zero_2017}. The angle of high absorption and emission can be modified by changing the film thickness, without altering the spectral peak position. Recently, broadband directional absorption and emission was demonstrated by layering polaritonic materials such that a gradient of ENZ layers was formed \cite{xu_broadband_2021, ying_whole_2022}. Building on this finding, superior design flexibility was demonstrated through an InAs-based gradient ENZ photonic structure\cite{hwang_simultaneous_2023}, where arbitrary control over the plasma frequency of each layer was achieved by changing the doping level spatially\cite{law_mid-infrared_2012,law_doped_2014}. However, in photolithography-free planar thin-films, the Berreman mode and all ENZ-based strategies for modifying thermal emission are inherently limited to transverse magnetic (TM) radiation. This is due to the fact that, for transverse electric (TE) waves, the electric field lies parallel to the surface and is unchanged across a boundary where only the dielectric permittivity changes.  

We first show that to achieve comparable control over TE waves, one must introduce a discontinuity in the magnetic permeability instead. However, in the visible, near-IR, and mid-IR spectral ranges, no known natural materials have magnetic activity \cite{landau2013electrodynamics}. Thus, magnetic properties must be engineered. A typical approach is to pattern structures with electrical resonances that induce a magnetic response, but this gets increasingly difficult as frequencies enter the THz regime \cite{monticone_quest_2014}. Recently, optical magnetism was demonstrated using layered metamaterials comprised of alternating dielectric and metallic layers \cite{papadakis_optical_2018}. Notably, the authors performed parameter retrieval \cite{papadakis_retrieval_2015} to demonstrate that the hyperbolic metamaterial had non-unity permeability, and demonstrated an EMNZ point in one of the structures. It should be noted that similar research on the interaction of transverse electric radiation with layered meta-materials has been referred to as the generalized Brewster angle effect\cite{watanabe_s-polarization_2008,sreekanth_generalized_2019}.

In this paper, we present a hyperbolic material comprised of alternating layers of doped and intrinsic InAs that shows spectral and directional control of thermal emission across both polarizations through a plasmon-based EMNZ response that is tunable over a broad spectrum. We theoretically demonstrate that that an MNZ film atop a perfect magnetic conductor absorbs TE radiation identically to how the Berreman mode absorbs TM radiation - calling this the "magnetic" Berreman mode. We then show that a hyperbolic metamaterial composed of alternating layers of doped- and undoped-InAs acts as a spectral and directional selective TE emitter, with an effective MNZ point. We use genetic and gradient descent optimization algorithms to arrive at a dual-polarized directionally and spectrally selective emitter. Parameter retrieval is performed to show that there is indeed a simultaneous EMNZ point near the absorption peak wavelength and angle. Finally, we fabricate the metamaterial stack and characterize its absorptivity.

\section{Results}

\begin{figure}[hbt]
\centering
\includegraphics[width=0.9\linewidth]{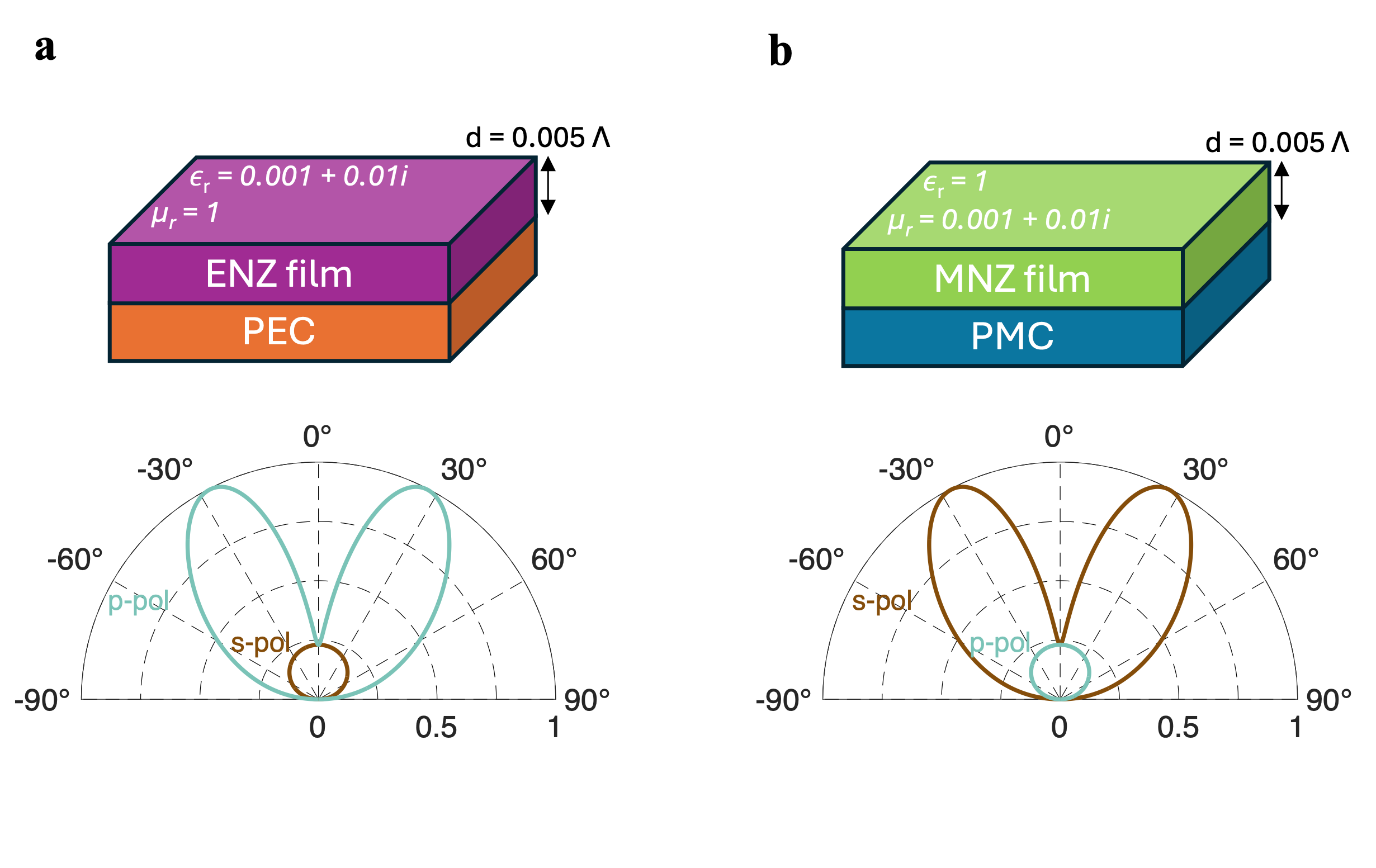}
\caption{Simulated emissivity, showing the directional and polarized emissivity of a (a) non-dispersive epsilon-near-zero film  ($\epsilon_r = 0.001+0.01i, \mu_r = 1$) with thickness $d=0.005~\Lambda$ on a perfect electric conductor substrate ($\epsilon_r = 1+100i, \mu_r = 1$). (b) a non-dispersive mu-near-zero film ($\mu_r = 0.001 + 0.01i, \epsilon_r = 1$) with thickness $d = 0.005~\Lambda$ on a perfect magnetic conductor substrate ($\epsilon_r = 1, \mu_r = 1+100i$).}
\label{fig:idealMNZ}
\end{figure}

Our approach to enabling lithography-free directional control of emissivity across both polarizations is to leverage and optimize an MNZ response enabled by a metamaterial-based approach. To motivate this strategy, we first demonstrate that a sub-wavelength, non-dispersive MNZ film on a perfect magnetic conductor (PMC) exhibits a TE absorption mode identical to the TM Berreman mode. Figure \ref{fig:idealMNZ} shows the results of a 1D transfer matrix method (TMM) simulation of the TM emissivity of a non-dispersive ENZ film with $\epsilon_r = 0.001 + 0.01i$ and $\mu_r = 1$ on a perfect electric conductor. The film thickness is sub-wavelength so that the Berreman mode may be excited \cite{campione_theory_2015} ($d = 0.005 \Lambda$, where $\Lambda$ is the mean incident wavelength). The results are consistent with past work showing strong broadband emission within a narrow angular range \cite{xu_broadband_2021}. There is no enhancement of TE emissivity (Figure \ref{fig:idealMNZ}b), and the unpolarized emissivity is therefore approximately halved (Figure \ref{fig:idealMNZ}c). Figures \ref{fig:idealMNZ}c - \ref{fig:idealMNZ}e show the simulated absorption response of a non-dispersive MNZ film on a perfect magnetic conductor, where the film properties are identical to the non-dispersive ENZ film, except that the permittivity and permeability values are exchanged. The result is that TM radiation is weakly absorbed whereas TE radiation exhibits strong broadband absorption. This transverse electric absorption mode can be thought of as a magnetic Berreman mode. 

\begin{figure}[hbt]
\centering
\includegraphics[width=0.9\linewidth]{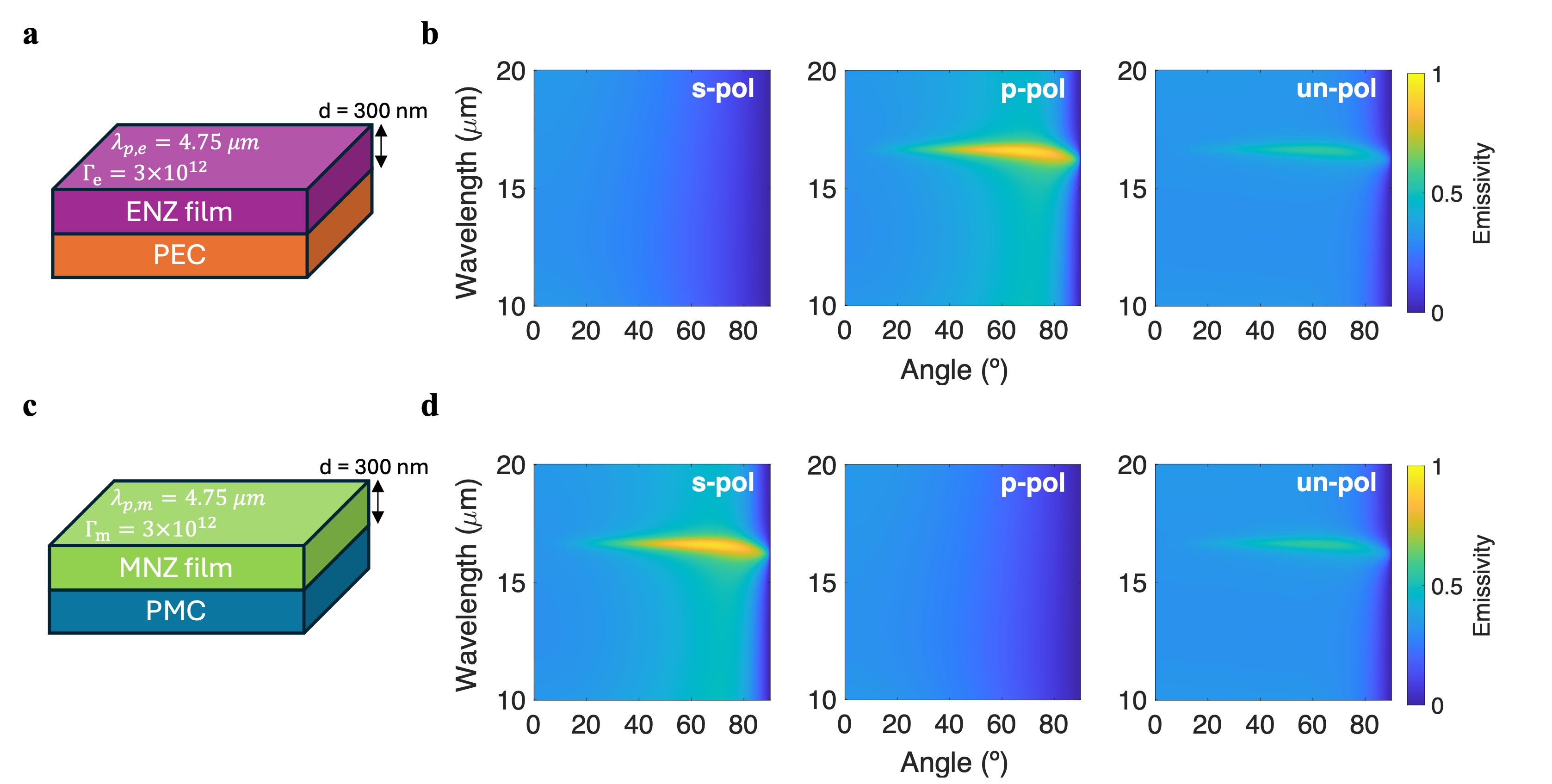}
\caption{ 
(a) The complex permittivity, permeability, impedance, and refractive index of a theoretical material with permittivity characterized by the Drude model ($\lambda_p = 15~\mu m, \Gamma = 0.06\omega_p~s^{-1}$) and unity permeability. 
(b) Simulated emissivity of a 300~nm film on a perfect electric conductor, revealing a high-angle TM emissivity peak identified as the Berreman mode. 
(c) The complex permittivity, permeability, impedance, and refractive index of an imaginary material with permeability defined by the Drude model ($\lambda_p = 15~\mu m, \Gamma = 0.06\omega_p~s^{-1}$) and unity permittivity. 
(d) Simulated emissivity of a 300~nm film on a perfect magnetic conductor, displaying a high-angle TE emissivity peak newly identified as the magnetic Berreman mode.}
\label{fig:drude}
\end{figure}

We extend our study of the magnetic Berreman mode to dispersive films, with permittivity or permeability defined by the Drude model:

\begin{equation}
\epsilon(\omega) = \epsilon_{s,\infty} - \frac{\omega_p^2}{\omega^2 + i\omega\Gamma}
\label{eq:drude}
\end{equation}

where $\omega_p=2\pi c/\lambda_p$ is the angular plasma frequency, $\Gamma = 1/\tau$ is the scattering rate, and $\epsilon_{s,\infty}$ is the high-frequency dielectric constant. Figure \ref{fig:drude}a-b shows the simulated optical properties of a plasmonic film with permittivity defined by the Drude model, featuring the characteristic transverse magnetic Berreman mode. Figure \ref{fig:drude}c-d shows the simulated optical properties of a hypothetical magnonic film with permeability defined by the Drude model. See Methods.

\begin{figure}[hbt]
\centering
\includegraphics[width=0.9\linewidth]{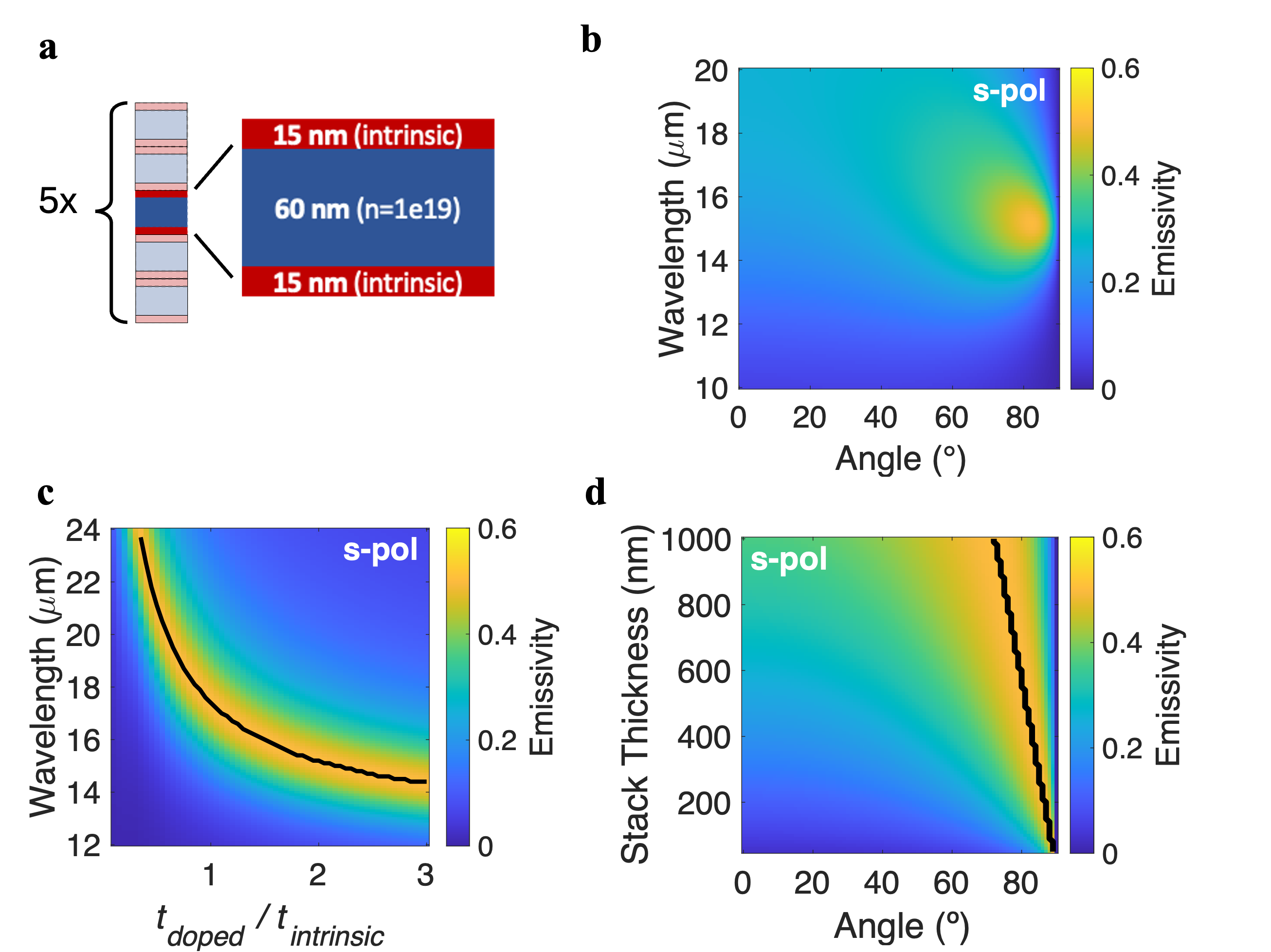}
\caption{
(a) An InAs-based metallodielectric metamaterial comprising five repeating units, each with 60 nm of n=1e19 InAs and 30 nm of intrinsic InAs, resulting in a total stack height of 450 nm and a $t_{doped}/t_{intrinsic}$ ratio of 2.
(b) Simulated emissivity revealing a high-angle transverse electric emissivity peak.
(c) TE spectral emissivity plotted against the fraction of doped-InAs to intrinsic-InAs while keeping the total stack thickness constant, revealing a decrease in peak-emissivity wavelength as the fraction increases.
(d) TE angular emissivity plotted against total stack thickness while maintaining a constant ratio of doped-InAs to intrinsic-InAs, demonstrating the shift of the peak-emission angle towards normal incidence as the total stack thickness increases.
}
\label{fig:MD}
\end{figure}

Our approach to an effective MNZ metamaterial is to combine the high-frequency magnetic response of metallodielectric hyperbolic metamaterials with the highly tunable plasmon resonance of doped III-V semiconductors. We choose InAs since it has a high-quality infrared plasmonic resonance owing to its small electron effective mass and its ability to incorporate a high density of dopants \cite{law_doped_2014}. Figure \ref{fig:MD}a shows a schematic of an InAs-based hyperbolic metamaterial comprised of doped-InAs ($n=1e19$) and intrinsic InAs. The stack is characterized by its total height and by the ratio of doped InAs to undoped InAs. Figure \ref{fig:MD}b shows the simulated emissivity, revealing a high-angle, transverse electric mode. Figure \ref{fig:MD}b shows that the proportion of doped-InAs ($n=1e19$) to undoped-InAs controls the spectral position of the TE absorption peak, demonstrating the true metamaterial nature of this film. Note that the peak width increases as the fraction decreases, and that the peak position asymptotically approaches the ENZ frequency of the doped-InAs constituent as the ratio increases. Figure \ref{fig:MD}f shows that increasing stack height moves the angular peak towards normal incidence, though the effect is weak. These properties are reminiscent of the plasmonic Berreman mode, where doping concentration modulates the spectral peak and film thickness modulates the angular peak.

We highlight that the TE emissivity is a genuine metamaterial effect: although the doping concentration (and thus plasma resonance) remains the same, the TE emission peak can shift by adjusting the ratio of doped to intrinsic semiconductor layer thicknesses. Notably, the TE response is not the direct result of the doped InAs layers, since doped InAs still has unity permeability; rather, the effective permeability arises from interference between layers. 

In addition we note that a single slab of doped InAs exhibits TE emissivity, with its peak positioned at the plasma frequency. This can be deduced from the trend in Figure \ref{fig:MD}c as the ratio of doped layer thickness to intrinsic layer thickness goes to infinity. Such a free-standing plasmonic film also supports a TM absorption mode at the same frequency, thus achieving dual-polarized, directional emission. However, this dual-polarized emissivity can only be achieved when the incident and substrate media are identical. In practice, where the incident medium differs from the substrate, the TE emission peak disappears. 

\begin{figure}[hbt]
\centering
\includegraphics[width=0.9\linewidth]{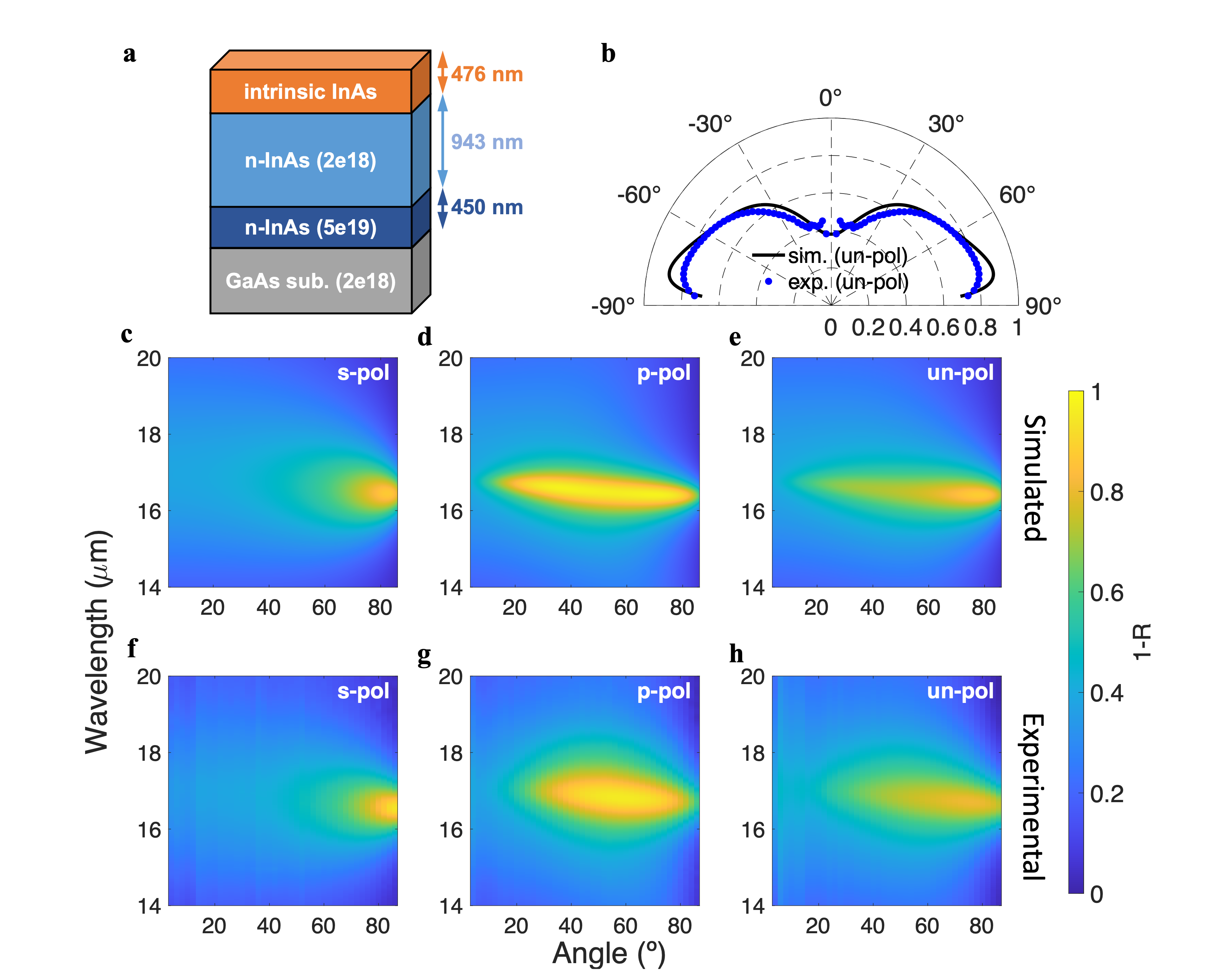}
\caption{
(a) Schematic representation of the InAs metamaterial showing optimized thicknesses and doping concentrations achieved through a combination of the genetic algorithm and gradient descent algorithm.
(b) A polar plot comparing the simulated unpolarized emissivity to the experimentally measured unpolarized emissivity at the peak-emissivity wavelength, highlighting high angular-emissivity contrast.
(c)-(e) Simulated s-, p-, and unpolarized emissivity of the optimized EMNZ metamaterial, demonstrating spectrally-aligned directional emissivity in both polarizations.
(f)-(h) Spectral-angular experimental measurements of the MBE-grown EMNZ metamaterial, validating the metamaterial's unpolarized directional emissivity.
}
\label{fig:experimental}
\end{figure}

We relax constraints on the number of layers, the doping concentration, and layer thickness and use a genetic algorithm and gradient descent optimizer to design a stack that aligns the spectral emission peaks of both polarizations and maximizes peak-to-normal angle emissivity contrast of unpolarized radiation. The result is illustrated in Figure \ref{fig:experimental}a. It is possible to design a similar stack with arbitrary control over the position of the TM and TE peaks independently. The TM spectral peak is primarily determined by the plasma frequency of the doped-semiconductor, and the TE peak is determined both by the plasma frequency of the doped-semiconductor and the fraction of doped- and intrinsic semiconductor. 

We experimentally demonstrate lithography-free directional thermal emission across both polarizations. The optimized stack is epitaxially grown on a single-side polished GaAs wafer [001] (n=2e18) using a VEECO Gen-930 Molecular Beam Epitaxy reactor (see Methods). 

\begin{figure}[hbt]
\centering
\includegraphics[width=0.9\linewidth]{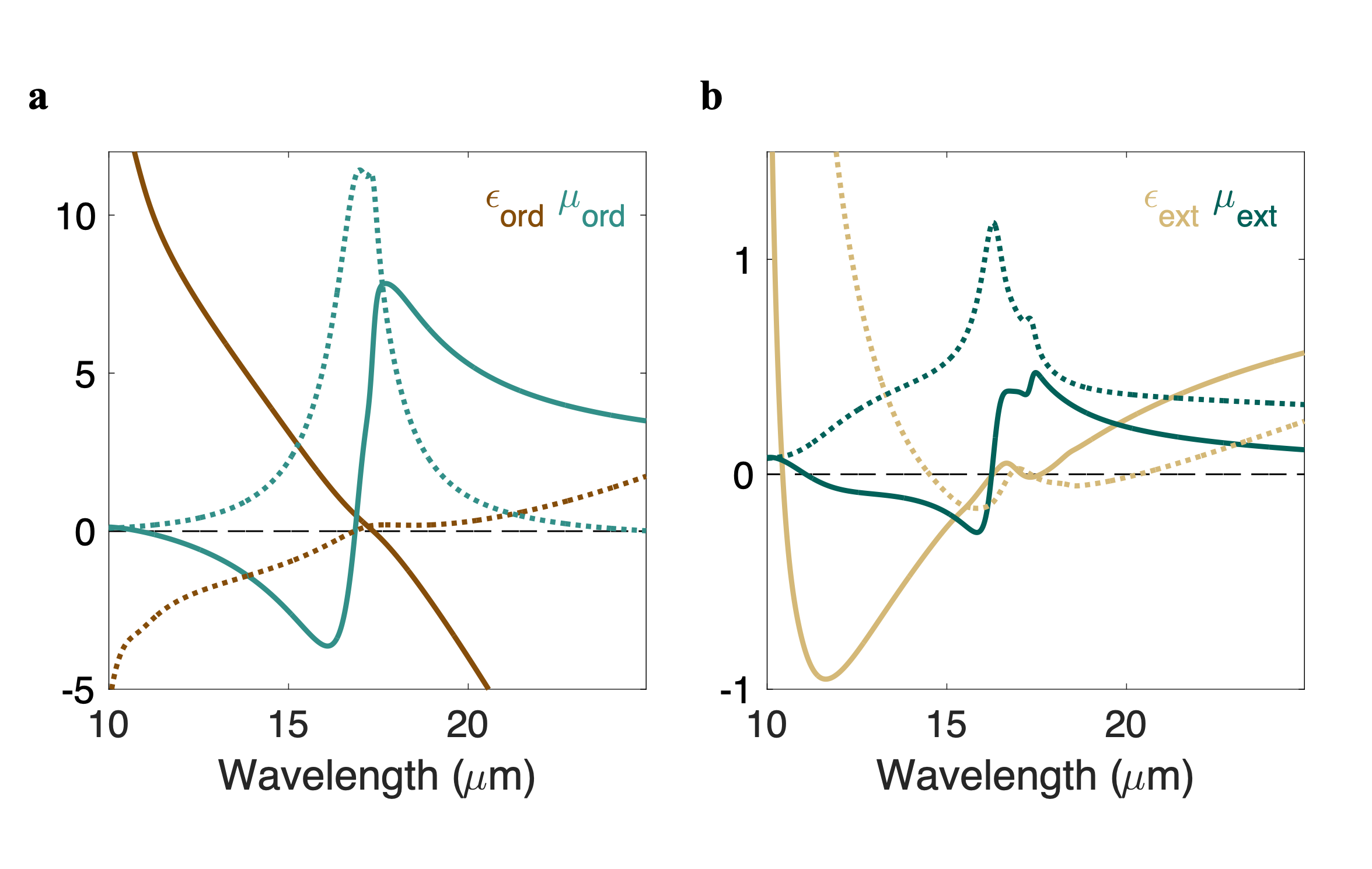}
\caption{
Retrieved parameters assuming a uniaxial media, providing confirmation that the metamaterial exhibits a spectral region in which both epsilon and mu are near-zero (EMNZ).
}
\label{fig:EMNZ_paramRetrieval}
\end{figure}

Figure \ref{fig:experimental}b shows a polar plot of the unpolarized emissivity comparing simulation to experimental. The stack exhibits an unprecedented maximum emissivity of 0.81 while sustaining a peak-to-normal incidence (4\textdegree) emissivity delta of 0.43. Figure \ref{fig:experimental}c-\ref{fig:experimental}e show the simulated emissivity of the optimum stack using Drude parameters extracted from the MBE calibration samples. Figure \ref{fig:experimental}f-\ref{fig:experimental}h show the experimentally measured emissivity showing great agreement with the simulation. The experimental and simulated emissivity characteristics are shown in Table \ref{tab:peak_characteristics}. Note that normal incidence is actually 4\textdegree, due to experimental limitations. Figure \ref{fig:EMNZ_paramRetrieval} shows the retrieved parameters of the metamaterial assuming uniaxial anisotropy. The stack exhibits EMNZ behavior with a simultaneous zero-crossing of the retrieved parameters about the design wavelength.

\begin{table}
  \centering
  \begin{tabular}{ccc}
    \toprule
    \textbf{Characteristics (unpolarized)} & \textbf{Simulated} & \textbf{Experimental} \\
    \midrule
    Peak Emissivity & 0.88 & 0.81 \\
    Peak Wavelength $(\mu m))$ & 16.4 & 16.7 \\
    Peak Angle $(^\circ)$ & 78.1 & 76.0 \\
    Peak-to-normal Emissivity Contrast & 0.50 & 0.43 \\
    Ratio (peak):(normal emissivity)  & 2.3:1 & 2.1:1 \\
    \bottomrule
  \end{tabular}
  \caption{Characterization of Unpolarized Emissivity Peak in EMNZ Metamaterial}
  \label{tab:peak_characteristics}
\end{table}

\begin{figure}[hbt]
\centering
\includegraphics[width=\linewidth]{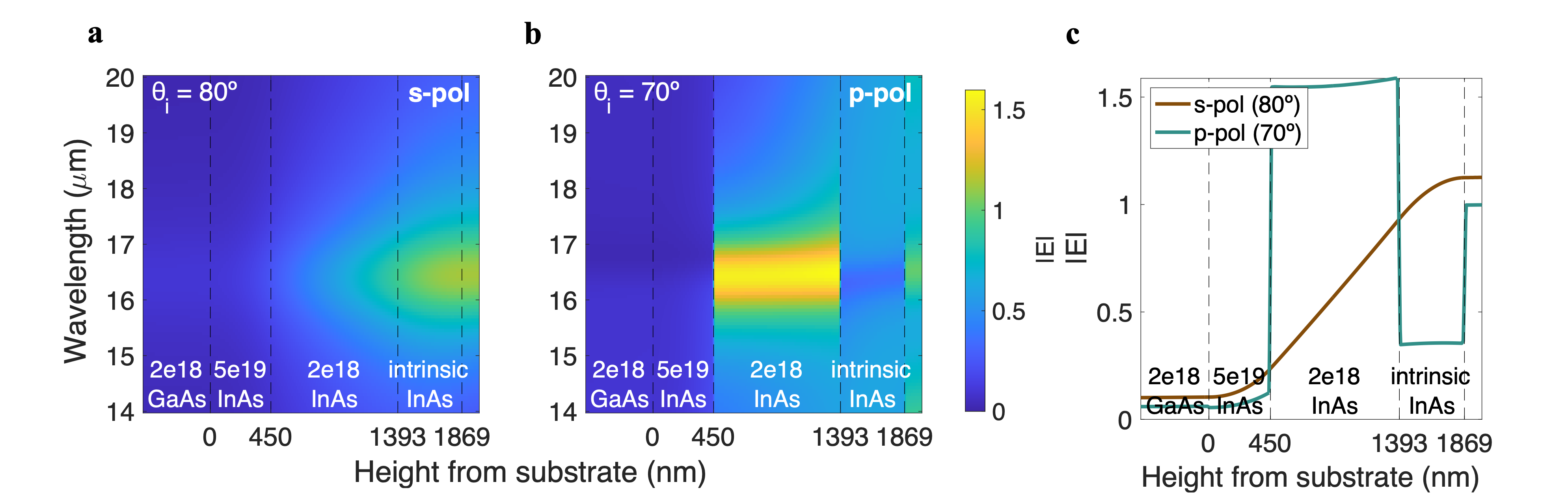}
\caption{
FDTD simulated spectral electric-field intensity profiles for (a) TE and (b) TM light within the EMNZ structure, each corresponding to the peak emission angle of that polarization.
(c) Electric field depth profiles at each polarization's peak emission wavelength and angle within the EMNZ structure.
}
\label{fig:efield}
\end{figure}

\section{Conclusion}

In this article, we present a method for achieving directional emission across both polarizations using a hyperbolic metamaterial composed of doped semiconductors. Theoretically, we demonstrate that an MNZ film on a perfect magnetic conductor hosts a TE absorption mode analogous to the TM Berreman mode, and we dub this the "magnetic" Berreman mode. 

We then demonstrate how an effective MNZ film can be experimentally realized via a doped-semiconductor-based plasmonic hyperbolic metamaterial. We enlist numerical optimization to arrive at a fabricable design that can be described as having an epsilon-and-mu-near-zero (EMNZ) point that drives directional unpolarized emission. This extends previous work on planar epsilon-near-zero metamaterials, which showed directional emission only for TM radiation. Our result shows an unprecedented, unpolarized maximum emissivity of 0.81 while sustaining a peak-to-normal incidence (4\textdegree) emissivity delta of 0.43.

\section{Methods}

\subsection{Simulation Details}
Figure \ref{fig:drude}a an isotropic material where the permittivity is defined by the Drude model with $\lambda_p=15~\mu m$ and $\Gamma = 0.06\omega_p~s^{-1}$. The permeability is set to unity. Figure \ref{fig:drude}b shows the emissivity of 300~nm of this material on a perfect electric conductor ($\epsilon_r = 1+100i, \mu_r = 1$), revealing the narrowband, high angle Berreman mode for TM radiation, consistent with a plasmonic film on a metal. Figure \ref{fig:drude}c shows the optical properties of a hypothetical magnetron plasmonic film, where the \textit{permeability} is defined by the Drude model with $\lambda_p=15~\mu m$ and $\Gamma = 0.06\omega_p~s^{-1}$. Note that the impedance and refractive index are identical. Finally, we enforce passivity when calculating the refractive index and impedance, ensuring both that the imaginary component of the refractive index is positive and that the real component of the impedance is positive.

\subsection{MBE Growth}
First, calibration samples consisting of single films of Si-doped InAs were epitaxially grown on a single-side polished GaAs film (n=2e18) using a VEECO Gen-930 Molecular Beam Epitaxy reactor. The wafer was mounted to an aluminum chuck using indium paste. A 100~nm GaAs buffer layer was grown. Two calibration wafers were grown consisting of 300~nm of 2e18 InAs and 5e19 InAs. FT-IR reflectivity data were collected using a Bruker Invenio-R system and a Drude model was fit to extract both the plasma frequency $\omega_p$ and the scattering time $\tau$, similar to methods used by other researchers \cite{li_infrared_1993,law_epitaxial_2013}. In addition, the free carrier concentration and carrier mobility were determined through Hall measurements made on a custom Van der Pauw set-up. The results are shown in Table \ref{tab:calibration_data}.

\begin{table}
  \centering
  \begin{tabular}{ccccc}
    \toprule
    \textbf{Nominal Doping \& Thickness} & \textbf{$\lambda_p~(\mu m)$} & \textbf{$\tau~(s)$} & \textbf{$n~(cm^{-3})$} & \textbf{$\mu_n~(\frac{cm^2}{Vs})$} \\
    \midrule
    2e18 (300~nm) & 16.8 & 2.7e-14 & 2.7e18 & 5330 \\
    5e19 (300~nm) & 6.2 & 3.0e-13 & 5.3e19 & 850 \\
    \bottomrule
  \end{tabular}
  \caption{Optical and Electrical Properties of MBE Calibration Samples}
  \label{tab:calibration_data}
\end{table}

The optimum design layer thicknesses were tuned after extracting the Drude parameters from the calibration films to maximize the unpolarized peak emissivity. We epitaxially grow the optimized and tuned stack on a three-inch single-side polished GaAs substrate doped 2e18. Infrared reflectivity data was collected with a Bruker Invenio-R system from $400~cm^{-1}$ to $8000~cm^{-1}$ in $4~cm^{-1}$ increments from 4º to 86º in 2º increments.

\section*{Acknowledgements}

This material was based upon work supported by the National Science Foundation under grant no. ECCS-2146577

\bibliographystyle{apsrev4-2}   
\bibliography{references_MNZ,references_manual}

\end{document}